\documentclass[%
 preprint,
 amsmath,amssymb,
 aps,
]{revtex4-2}

\usepackage{graphicx}
\usepackage{dcolumn}
\usepackage{bm}

\begin{document}

\preprint{arXiv}

\title{Joule-Class THz Pulses from Microchannel Targets}

\author{G. Bruhaug\textsuperscript{1,*},H.G. Rinderknecht\textsuperscript{1}, K. Weichman\textsuperscript{1}, M.VanDusen-Gross\textsuperscript{1},J.P. Palastro\textsuperscript{1}, M.S. Wei\textsuperscript{1}, S.P. Regan\textsuperscript{1}, Y. E\textsuperscript{2}, K. Garriga\textsuperscript{2}, X.-C. Zhang\textsuperscript{1,2}, G.W. Collins\textsuperscript{1}, J.R. Rygg\textsuperscript{1}}
\affiliation{\textsuperscript{1}Laboratory for Laser Energetics, University of Rochester, Rochester, NY 14623-1299, USA}
\affiliation{\textsuperscript{2}The Institute of Optics, University of Rochester, Rochester, NY, 14627, USA}

\affiliation{\textsuperscript{*}gbruhaug@ur.rochester.edu}

\date{\today}

\begin{abstract}
\textbf{Inference of joule-class THz radiation sources from microchannel targets driven with hundreds of joule, picosecond lasers is reported. THz sources of this magnitude are useful for nonlinear pumping of matter and for charged-particle acceleration and manipulation. Microchannel targets demonstrate increased laser-THz conversion efficiency compared to planar foil targets, with laser energy to THz energy conversion up to $\sim$0.9\% in the best cases.}
\end{abstract}

\maketitle


In this Letter we present the first experimental evidence of joule-class terahertz pulses from microchannel targets driven by picosecond-class lasers. These measurements are made from a single point of measurement and then the total THz emission estimated via a model of THz production during the laser-microchannel irradiation. Microchannel targets irradiated by picosecond class lasers show an increased laser-THz conversion efficiency, which is believed to be due to an increased laser-electron conversion efficiency. 
\\
THz radiation occupies the range between microwave and infrared radiation and has been historically difficult to generate at high powers and pulse energies \cite{Liao2019,Hafez2016IntenseApplications,Herzer2018}. Typical laboratory and industrial THz sources operate in the tens of mW range for temperature limited continuous sources \cite{Khalatpour2021High-powerSystems} and $\mu$J range for pulsed sources\cite{Liao2019,Hafez2016IntenseApplications,Herzer2018} as limited by material damage or laser drive energy. Exceeding tens of mJ pulse energies and GW's of peak THz power, opens up new avenues for study of THz-matter interactions and has been a goal of recent research \cite{Liao2020,Herzer2018,Lei2022HighlyInteractions,Zeng2020}. High-energy, high-power THz irradiation can drive matter between phases (such as between insulating and conductive) \cite{Fulop2020Laser-DrivenSources}, be used to probe the low-frequency limit of relativistic light matter interactions, and provide alternate paths for acceleration and manipulation of charged particles\cite{Salen2019MatterTechnology,Hamm2017Perspective:Molecules,Hafez2016IntenseApplications,Fulop2020Laser-DrivenSources,Zhang2021IntenseApplication,Zhang2017ExtremeScience}. 

THz radiation can be generated via a wide variety of methods, with laser-based production being the most common. Typical laboratory and industrial THz sources utilize laser irradiation of nonlinear crystals to generate THz pulses up to single mJ's of pulse energy \cite{Zhang20211.4-mJNiobates} (limited by crystal damage), or utilize quantum cascade lasers to generate tens of mW CW sources \cite{Khalatpour2021High-powerSystems}. Higher THz pulse energies and peak powers are achievable using high-power laser interactions with gases \cite{Dechard2018TerahertzRadiation}, liquids \cite{E2021BroadbandLiquids}, and solids \cite{Liao2020,Lei2022HighlyInteractions,Zeng2020,Herzer2018, Zhang2022,Zhuo2017}. Laser solid interactions have generated the highest peak power and energy THz sources to date \cite{Liao2019,Liao2020} via coherent transition radiation (CTR) from laser generated electrons. Previous work with $\sim$$10^{19}$ W/cm\textsuperscript{2}, $>$60-J laser irradiation of metal foil targets have produced up to $\sim$200-mJ THz yields (corresponding to $>$100-GW peak power) \cite{Liao2019,Liao2020}. 

Further increases in the THz energy and power can be obtained using structured targets such as microchannels. This is inline with the over-all trends of THz research to focus on structures on the order or smaller of the wavelength of THz radiation \cite{Kumar2022PhotoinducedParameters,Kang2024FrequencyMetamaterials}. Microchannel targets consist of $\mu$m-scale-diameter, hundreds-of-$\mu$m-long channels in a solid substrate that act as waveguides for the laser pulses \cite{Rinderknecht2021,Yi2019, Rosmej2019InteractionGamma-rays}. Coupling a >$10^{19}$ W/cm\textsuperscript{2} laser pulse into these waveguides can produce a relativistically transparent plasma by blowdown of the channel walls. The laser is constrained to interact with this plasma at high intensity for many Rayleigh lengths, generating extremely large currents and magnetic fields. These large currents and fields are predicted to be efficient sources of gamma rays \cite{Rosmej2019InteractionGamma-rays} and THz radiation \cite{Yi2019} when driven by tens of femtosecond laser pulses. However, the interaction of picosecond-class laser pulses with microchannel targets has remained relatively unexplored until now. There also remains the potential to tune the output radiation of the microchannel by altering the driving laser pulse length and intensity, the microchannel size, and the microchannel fill density \cite{Rinderknecht2021,Yi2019}.

Experiments were performed on the OMEGA EP laser at the Laboratory for Laser Energetics to investigate THz production by picosecond-duration laser-microchannel interaction as shown in Fig. 1. This laser can only provide 14 shots per day, which limits the data collection capability but is partially compensated by the high laser energy and intensity available. The 1054-nm central wavelength OMEGA EP laser was used with a pulse duration of $0.7 {\pm} 0.1$ ps. Laser pulse energies ranged from $\sim$150 to $\sim$300 J. The laser spot size (80\% of laser energy contained in this radius) is $\sim$30 $\mu$m at best focus, which provides an average irradiance of $\sim$$10^{19}$ W/cm\textsuperscript{2} on target. 
Although the average intensity is only moderately relativistic, the peak intensity is substantially higher ($>10^{20}$ W/cm\textsuperscript{2}), due to the complex structure of the OMEGA EP laser spot [Fig. 1(c)].

\begin{figure}
\centering
\includegraphics[width=\linewidth]{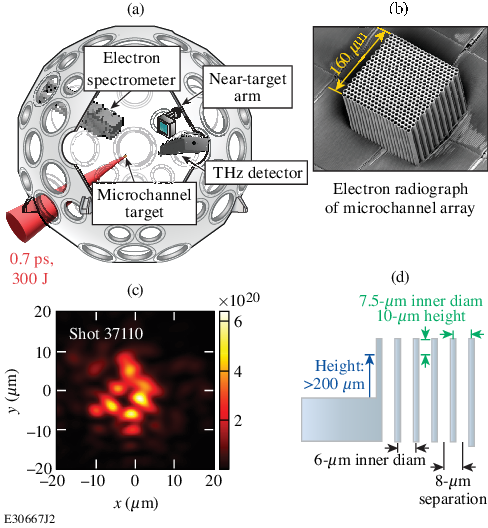}
\caption{ (a) A diagram of the location of the detectors relative to the laser beam and target. A 2-dimensional projection of this figure can be seen in the Supplemental Materials. (b) An electron microscopy image of one of microchannel targets used, (c) an example laser spot from OMEGA EP showing the greater peak irradiance than $10^{19}$ W/cm\textsuperscript{2} with transverse $x$ and $y$ laser spot dimensions at best focus shown, and (d) a cross section of the microchannel targets used. }
\end{figure}

The targets used in these experiments were 6-$\mu$m inner diameter, 8-$\mu$m channel-to-channel separation microchannel arrays produced in a CH substrate (polyimide film specifically) via two-photon polymerization and mounted onto SiC stalks. The arrays varied in length from 40 $\mu$m to 400 $\mu$m. Both empty channels and channels filled with 5 or 8 mg/cm\textsuperscript{3} of CH foam were used, but no trends with channel fill were seen in these experiments and as such no distinction in channel fill will be made in reporting microchannel results. The microchannels are stable at room temperature and pressure, which makes handling and positioning of the targets much simpler. The choice of CH (and specifically polyimide) was driven by target construction constraints. Further information on the construction of these targets can be found in reference \cite{Rinderknecht2021}. A picture and diagram of a typical microchannel array can be seen in Fig. 1(b) and 1(d) respectively. To provide a standard THz baseline from the OMEGA EP laser, 20-$\mu$m thick Cu foils were used. These also act as a comparison to previous laser-foil experiments at other laser facilities \cite{Herzer2018,Liao2019,Liao2020}.

For these experiments a single THz background/energy meter (TBEM) \cite{Bruhaug2022DevelopmentLaser} detector was positioned nearly orthogonal to the laser axis and 167.5 cm from the target, which is a nearly ideal location based on previously reported THz emission patterns from foils \cite{Liao2019,Herzer2018}. The TBEM detector utilized a 10.9 THz low pass filter, silicon wafer, and polytetrafluoroethylene pyrometer window to eliminate all infrared, optical and ultraviolet light \cite{Bruhaug2022DevelopmentLaser}. These filters limit the detectable THz frequency range from 0.1 to 10.9 THz but provided no other spectral information. The electron diagnostics for the experiment were an Electron positron proton spectrometer (EPPS) \cite{Ayers2010} covering 1 to 200 MeV in energy and a stack of radiochromic film positioned with a near target arm (NTA), to study  >1-MeV electron emission patterns. Due to the nearly perpendicular position of the EPPS, the resulting measurements are not wholly indicative of the electron spectrum responsible for the THz emission. A diagram showing the placement of the detectors relative to the laser beam and target can be seen in Fig. 1(a).

We infer the THz emission fraction intercepted by the detector by integrating Eq. (1) over all detectable THz frequencies and the area and angles covered by the THz detector shown in Fig. 1(a) and dividing that value by an integration over all emission angles to determine the fraction of THz detected. More information about the inference of the THz pulse energy can be found in the Supplemental Materials. The THz pulse energy inferences vs laser energy and intensity are shown in Fig. 2(a) and 2(b), respectively. The laser-THz conversion efficiency is then plotted vs laser energy and intensity in Fig. 2(c) and 2(d) respectively. The error bars are generated via the known error of the TBEM detectors \cite{Bruhaug2022DevelopmentLaser} added in quadrature with variances in the CTR model for laser pulse length, electron temperature, and emission profile. Theoretical curves are added to the plot for various laser-to-electron conversion efficiencies using Equations (2) through (6).

\begin{figure}
\centering
\includegraphics[width=\linewidth]{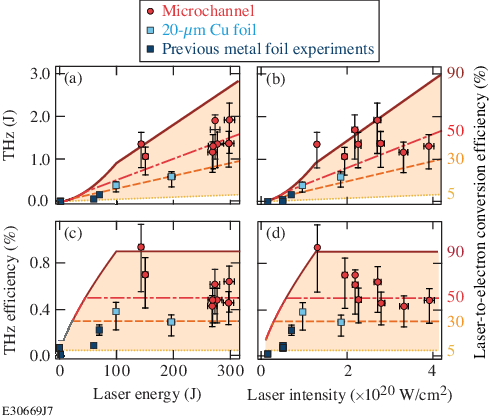}
\caption{(a) THz yield vs laser pulse energy and (b) laser pulse irradiance as well as the (c) laser-to-THz generation efficiency vs laser pulse energy and (d) laser pulse irradiance for microchannel targets on OMEGA EP compared to previous laser-solid-THz experiments on OMEGA EP and other facilities\cite{Liao2016,Liao2019,Liao2020,Herzer2018,Gopal2013CharacterizationInteraction}. Theoretical THz emission trend lines are added for laser-to-electron conversion efficiencies of $\eta =$ 5\%, 30\%, 50\%, and 90\% and CTR clamp factor $\alpha =$ 0.1.}.
\end{figure}

Microchannel targets (red points in Fig. 2) were inferred to generate joule-scale THz pulses [Fig. 2(a) and 2(b)] at laser-THz efficiencies reaching up to $\sim$0.9\% [Fig. 2(c) and 2(d)]. The <1 ps pulse of the laser, implies a $\sim$1 ps THz pulse duration \cite{Liao2019} and in turn implies peak powers in the terrawatt regime. For the same laser pulse irradiance and energy, the microchannel targets are inferred to yield nearly a factor of 10 more THz energy than the flat foil targets and are a factor of 2 to 3 more efficient at THz production in the best cases. This larger THz output likely originates from a higher laser-to-electron conversion efficiency, although further experiments will be needed to confirm this hypothesis. The microchannel targets are most consistent with 50\% laser-to-electron conversion efficiency, while the foil targets lie near the 30\% line or lower. The later is consistent with previously reported laser-to-electron conversion efficiencies for flat metal foils \cite{Nilson2011ScalingInteractions, Myatt2007High-intensityEP, Roth2016}.

There was a wide variation (factor of 2) in microchannel laser-THz efficiency, however and no clear trends within the microchannel shots with irradiance [Fig. 2(c) and 2(d)]. This variability may be attributable to shot-to-shot changes in the OMEGA EP beam pointing and intensity profile. When the most-intense points of the laser pulse happen to align with one of the microchannels in the array, many more relativistic electrons are generated and able to radiate THz radiation. This best-case scenario does not always happen and as such, the performance of these microchannel targets can vary greatly shot to shot. Further refinement of CTR measurements and target modeling may allow for indirect measurement of the total bunch charge transiting the target surface, although this calculation must ultimately account for the reflection of electrons by sheath fields \cite{Denoual2023ModelingPulses}.

 The theoretical curves present in Fig. 2 are generated via a model built off of previous work on CTR THz and laser-electron generation efficiency. The typical relativistic-intensity laser solid THz experiment relies primarily on coherent transition radiation (CTR) as the THz generation mechanism\cite{Herzer2018,Liao2020,Liao2023PerspectivesPlasmas}. Although more complex targets may allow for other generation mechanisms, the large number of laser ejected electrons will always result in CTR emission as well \cite{Herzer2018,Liao2020,Liao2023PerspectivesPlasmas}. We believe that this is the dominant form of THz radiation generation in microchannel targets when irradiated by picosecond lasers as well, primarily because the more-complex generation mechanisms rely on tens of femtosecond laser-plasma physics \cite{Yi2019,Rinderknecht2021}. However, more efficient laser-electron generation does not require fs laser pulses and would increase the CTR yield dramatically. CTR is created by electron bunches transiting between the plasma and vacuum surface. The angular radiated CTR energy can be described with the Ginzburg Frank formula \cite{Liao2019,Schroeder2004,Herzer2018}. 

\begin{equation}
\begin{split}
    \frac{dW^2}{d\Omega d\omega} = \frac{e^2}{4\pi^3\epsilon_{0} c}N(N-1)\sin^{2}{\theta} \times \\ \Big|\int du \ g_{||}(u) F(\omega , \theta , u) \frac{u(1+u^{2})^{1/2}}{1+u^{2}\sin^{2}{\theta}} D(\omega , \rho , u , \theta) \Big|^{2}
\end{split}
\end{equation}.

Here \emph{N} is the number of electrons in the bunch, \emph{c} is the speed of light, $\epsilon_{0}$ is the electric permittivity constant, \emph{u} = $\gamma \beta$  is the normalized momentum of the electrons, and $\theta$ is the angle of emission \cite{Schroeder2004}. The equations $g_{|| (u)}$, $F(\omega , \theta , u)$, and $D(\omega , \rho , u , \theta)$ are the electron temperature, spatial, and diffraction adjustments respectively, to the CTR and $\rho$ is the spatial size of the target for diffraction adjustment of the emission. This equation assumes a cylindrically symmetric, low-divergence electron emission, which is the standard approximation used for other laser-foil CTR experiments \cite{Herzer2018,Liao2019}. Recent theoretical work has also investigated electron reflux and the resulting synchrotron and secondary CTR emission. Simulations find this to be a potentially large portion of the emitted energy in <100-fs laser irradiation, but to not change the angular emission profile substantially \cite{Denoual2023ModelingPulses}.

CTR emission is coherent up to the electron pulse duration \cite{Schroeder2004}, which typically corresponds to the laser pulse duration \cite{Liao2019}. For picosecond-class lasers, the coherent frequency range extends into the THz. The emitted radiation follows a broadband, Boltzmann-like distribution \cite{Schroeder2004,Liao2020,Herzer2018}. If Eq. (1) is integrated over all coherent wavelengths and over all angles (and the electron $\gamma$ is assumed to be $\gg1$), the total radiated THz energy from a bunch can be calculated via \cite{Schroeder2004}

\begin{equation}
    W_{CTR} \simeq (4 r_{e} m_{e} c^{2})N^2 \ln{\gamma}/\lambda_{min}
\end{equation}.

Here $r_{e}$ is the classic electron radius, $m_{e}$ is the mass of the electron, $\gamma$ is the Lorentz gamma factor for the electrons, and $\lambda_{min}$ is the minimum coherent radiation wavelength

\begin{equation}
    \lambda_{min} = \beta^{-1} c \tau_{laser}
\end{equation},

where $\beta$ is the normalized velocity of the electrons and $\tau_{laser}$ is the laser pulse length. In the case of this experiment, $\lambda_{min}$ is 210 ${\pm}$ 30 $\mu$m, based on the $0.7 {\pm} 0.1$ ps laser pulse length. Equation (2) is derived for monoenergetic bunches, but is relatively unaffected by larger energy distributions due to the logarithmic scaling with electron $\gamma$. The derivation assumes that the electrons are relativistic ($\gamma >> 1$) \cite{Schroeder2004}, which is not always true in laser-plasma experiments and may overestimate the THz production as a result. Consequently, targets that can generate highly relativistic electrons are better suited for THz generation due to the logarithmic scaling with electron $\gamma$.

Equations (1) and (2) show that the total charge of the electron bunch driving the CTR process has a dramatic effect on the amount of THz radiation generated. Incoherent radiation scales linearly with charge compared to the quadratic scaling of coherent radiation, and can be neglected as a source of THz radiation in these conditions \cite{Schroeder2004,Liao2019,Herzer2018}.The strong scaling of THz emission with charge and the predicted high laser-to-hot-electron coupling efficiency motivates an investigation of microchannel targets as high-yield THz sources \cite{Rinderknecht2021,Yi2019}. 

A model for the laser-to-THz conversion efficiency can be constructed starting from the laser-to-hot-electron conversion efficiency $\eta$ \cite{Roth2016}, defined as

\begin{equation}
    \eta \simeq 0.1377 I_{19}^{0.74}, \quad \eta \leq \eta_{max}    
\end{equation}.

Here $I_{19}$ is irradiance in $ 10^{19}$ W/cm\textsuperscript{2}. This value ranges from $\eta<10\%$ \cite{Nilson2011ScalingInteractions,Myatt2007High-intensityEP} of the laser energy up to 90\% in some cases \cite{Roth2016,Ping2008AbsorptionRegime}. We assume the mean electron energy to be given by ponderomotive scaling\cite{Hatchett2000ElectronTargets}, which is used to estimate the THz energy reported.

\begin{equation}
    \varepsilon_{mean} \approx T(MeV) \approx \sqrt{I_{19}\lambda^{2}}
\end{equation}.

The laser wavelength is $\lambda$ in microns. Equations (4) and (5) can then be used to determine the total number of electrons for a given laser energy $E_L$ and irradiance as:  $N = E_{L} \eta /\varepsilon_{mean}$. The mean electron energy scales as the square root of laser irradiance while laser-to-electron efficiency scales almost linearly with irradiance up to $\eta_{max}$. Upon reaching that limit, the total number of generated electrons will fall with irradiance for a fixed laser energy, while the total energy in the electron bunch remains the same but the average energy per electron climbs.
Since CTR emission scales quadratically with the number of electrons while the total energy in the electron pulse scales linearly with the number of electrons, a limit must be put on the total CTR energy that can be emitted by the laser-generated electrons to preserve conservation of energy \cite{Schroeder2004,Krainara2018StudySource}. We will define a variable $\alpha$ as this limiting factor. This limiting factor represents the effect of the CTR emission on the electron beam, which will also induce self-limiting behavior that must be accounted for. The emission of CTR will slow the electron beam and spread the bunch, which in turn will change the coherency condition as the emission is occurs \cite{Krainara2018StudySource}. A fully self-consistent solution would require simulations of each particular case, but using Equations (2) through (5), simple scaling relations can be generated. To clamp the total CTR emitted from the electron bunch one can set the energy in the THz pulse equal to the energy in the electron bunch, solve for the total number of electrons, and multiply by $\alpha$ to clamp the CTR emission per electron, resulting in the following formula:

\begin{equation}
    N \approx \alpha\sqrt{\frac{\eta E_{laser} \lambda_{min}}{4m_{e}r_{e}c^{2}\ln{\gamma}}}
\end{equation}.

The $\alpha$ factor is assumed to be small enough to justify claiming that CTR is a secondary effect on the electron bunch. For the theoretical lines shown in Fig. 2, setting $\alpha = 0.1$ was found to fit the reported microchannel data along with as previous laser solid experiments \cite{Liao2019,Liao2020,Herzer2018,Liao2016,Gopal2013CharacterizationInteraction} best. Previous experiments using relativistic, monoenergetic, sub-nC electrons beams indicate that $\alpha \leq 0.3$ is an experimentally valid value \cite{Casalbuoni2009, DiMitri2018CoherentWakefield, Boscolo2009DESIGNLINAC}, which provides some credence to choosing a low value for $\alpha$. The wide angle of electron emission may further limit the ability of the laser-electron bunch to generate THz as compared to nearly a perfect accelerator generated electron bunch \cite{Schroeder2004}. The total THz emission can then be estimated for a given experiment using Equations (2) through (6) using assumptions of laser-to-electron conversion efficiency and electron-THz emission fraction and constitute the theoretical lines shown in Fig 2.

The simple model shown here motivates the hypothesis that microchannel targets driven by ps laser pulses are primarily efficient converters of laser-electron energy, which in turn will radiate more energy in the THz regime due to the $N^{2}$ scaling of CTR emission. The best performing microchannel targets (in red in Fig. 2) are well above the 50\% laser-electron conversion efficiency theoretical lines and are roughly double the laser-THz conversion efficiency of comparable foil targets shot with OMEGA EP laser (light blue points in Fig. 2). Even with the large variation in microchannel performance shot-to-shot, these targets consistently were inferred to generate joule-class THz pulses while no foils targets were able to reach those THz pulse energies. This could be indicative of new laser-THz generation mechanisms, but is more simply explained by higher laser-electron generation efficiency that then radiates THz via the well described CTR mechanism.

In conclusion, this work indicates that microchannel targets irradiated by hundreds of joule, picosecond-class lasers are efficient, energetic, and powerful sources of THz radiation. Joule-class THz yields are consistently inferred from microchannel targets using the OMEGA EP laser, motivating their use as future THz sources for extreme light matter experiments. Future work will seek to characterize both the spatial and spectral performance of this unique THz source as well as test new microchannel target designs to further optimize the source.

\begin{acknowledgments} This material is based upon the work supported by the Department of Energy, the National Nuclear Security Administration, under Award No. DE-NA0003856, Air Force Office of Scientific Research (FA9550-21-1-0300, FA9550-21-1-0389); National Science Foundation (ECCS-1916068), the University of Rochester, and the New York State Energy Research and Development Authority.
\end{acknowledgments}

\bibliography{apssamp}

\begin{thebibliography}{38}%
\makeatletter
\providecommand \@ifxundefined [1]{%
 \@ifx{#1\undefined}
}%
\providecommand \@ifnum [1]{%
 \ifnum #1\expandafter \@firstoftwo
 \else \expandafter \@secondoftwo
 \fi
}%
\providecommand \@ifx [1]{%
 \ifx #1\expandafter \@firstoftwo
 \else \expandafter \@secondoftwo
 \fi
}%
\providecommand \natexlab [1]{#1}%
\providecommand \enquote  [1]{``#1''}%
\providecommand \bibnamefont  [1]{#1}%
\providecommand \bibfnamefont [1]{#1}%
\providecommand \citenamefont [1]{#1}%
\providecommand \href@noop [0]{\@secondoftwo}%
\providecommand \href [0]{\begingroup \@sanitize@url \@href}%
\providecommand \@href[1]{\@@startlink{#1}\@@href}%
\providecommand \@@href[1]{\endgroup#1\@@endlink}%
\providecommand \@sanitize@url [0]{\catcode `\\12\catcode `\$12\catcode `\&12\catcode `\#12\catcode `\^12\catcode `\_12\catcode `\%12\relax}%
\providecommand \@@startlink[1]{}%
\providecommand \@@endlink[0]{}%
\providecommand \url  [0]{\begingroup\@sanitize@url \@url }%
\providecommand \@url [1]{\endgroup\@href {#1}{\urlprefix }}%
\providecommand \urlprefix  [0]{URL }%
\providecommand \Eprint [0]{\href }%
\providecommand \doibase [0]{https://doi.org/}%
\providecommand \selectlanguage [0]{\@gobble}%
\providecommand \bibinfo  [0]{\@secondoftwo}%
\providecommand \bibfield  [0]{\@secondoftwo}%
\providecommand \translation [1]{[#1]}%
\providecommand \BibitemOpen [0]{}%
\providecommand \bibitemStop [0]{}%
\providecommand \bibitemNoStop [0]{.\EOS\space}%
\providecommand \EOS [0]{\spacefactor3000\relax}%
\providecommand \BibitemShut  [1]{\csname bibitem#1\endcsname}%
\let\auto@bib@innerbib\@empty
\bibitem [{\citenamefont {Liao}\ \emph {et~al.}(2019)\citenamefont {Liao}, \citenamefont {Li}, \citenamefont {Liu}, \citenamefont {Scott}, \citenamefont {Neely}, \citenamefont {Zhang}, \citenamefont {Zhu}, \citenamefont {Zhang}, \citenamefont {Armstrong}, \citenamefont {Zemaityte}, \citenamefont {Bradford}, \citenamefont {Huggard}, \citenamefont {Rusby}, \citenamefont {McKenna}, \citenamefont {Brenner}, \citenamefont {Woolsey}, \citenamefont {Wang}, \citenamefont {Sheng},\ and\ \citenamefont {Zhang}}]{Liao2019}%
  \BibitemOpen
  \bibfield  {author} {\bibinfo {author} {\bibfnamefont {G.}~\bibnamefont {Liao}}, \bibinfo {author} {\bibfnamefont {Y.}~\bibnamefont {Li}}, \bibinfo {author} {\bibfnamefont {H.}~\bibnamefont {Liu}}, \bibinfo {author} {\bibfnamefont {G.~G.}\ \bibnamefont {Scott}}, \bibinfo {author} {\bibfnamefont {D.}~\bibnamefont {Neely}}, \bibinfo {author} {\bibfnamefont {Y.}~\bibnamefont {Zhang}}, \bibinfo {author} {\bibfnamefont {B.}~\bibnamefont {Zhu}}, \bibinfo {author} {\bibfnamefont {Z.}~\bibnamefont {Zhang}}, \bibinfo {author} {\bibfnamefont {C.}~\bibnamefont {Armstrong}}, \bibinfo {author} {\bibfnamefont {E.}~\bibnamefont {Zemaityte}}, \bibinfo {author} {\bibfnamefont {P.}~\bibnamefont {Bradford}}, \bibinfo {author} {\bibfnamefont {P.~G.}\ \bibnamefont {Huggard}}, \bibinfo {author} {\bibfnamefont {D.~R.}\ \bibnamefont {Rusby}}, \bibinfo {author} {\bibfnamefont {P.}~\bibnamefont {McKenna}}, \bibinfo {author} {\bibfnamefont {C.~M.}\ \bibnamefont {Brenner}}, \bibinfo {author} {\bibfnamefont {N.~C.}\ \bibnamefont
  {Woolsey}}, \bibinfo {author} {\bibfnamefont {W.}~\bibnamefont {Wang}}, \bibinfo {author} {\bibfnamefont {Z.}~\bibnamefont {Sheng}},\ and\ \bibinfo {author} {\bibfnamefont {J.}~\bibnamefont {Zhang}},\ }\bibfield  {title} {\bibinfo {title} {{Multimillijoule coherent terahertz bursts from picosecond laser-irradiated metal foils}},\ }\href {https://doi.org/10.1073/pnas.1815256116} {\bibfield  {journal} {\bibinfo  {journal} {Proceedings of the National Academy of Sciences of the United States of America}\ }\textbf {\bibinfo {volume} {116}},\ \bibinfo {pages} {3994} (\bibinfo {year} {2019})}\BibitemShut {NoStop}%
\bibitem [{\citenamefont {Hafez}\ \emph {et~al.}(2016)\citenamefont {Hafez}, \citenamefont {Chai}, \citenamefont {Ibrahim}, \citenamefont {Mondal}, \citenamefont {F{\'{e}}rachou}, \citenamefont {Ropagnol},\ and\ \citenamefont {Ozaki}}]{Hafez2016IntenseApplications}%
  \BibitemOpen
  \bibfield  {author} {\bibinfo {author} {\bibfnamefont {H.~A.}\ \bibnamefont {Hafez}}, \bibinfo {author} {\bibfnamefont {X.}~\bibnamefont {Chai}}, \bibinfo {author} {\bibfnamefont {A.}~\bibnamefont {Ibrahim}}, \bibinfo {author} {\bibfnamefont {S.}~\bibnamefont {Mondal}}, \bibinfo {author} {\bibfnamefont {D.}~\bibnamefont {F{\'{e}}rachou}}, \bibinfo {author} {\bibfnamefont {X.}~\bibnamefont {Ropagnol}},\ and\ \bibinfo {author} {\bibfnamefont {T.}~\bibnamefont {Ozaki}},\ }\bibfield  {title} {\bibinfo {title} {{Intense terahertz radiation and their applications}},\ }\bibfield  {journal} {\bibinfo  {journal} {Journal of Optics (United Kingdom)}\ }\textbf {\bibinfo {volume} {18}},\ \href {https://doi.org/10.1088/2040-8978/18/9/093004} {10.1088/2040-8978/18/9/093004} (\bibinfo {year} {2016})\BibitemShut {NoStop}%
\bibitem [{\citenamefont {Herzer}\ \emph {et~al.}(2018)\citenamefont {Herzer}, \citenamefont {Woldegeorgis}, \citenamefont {Polz}, \citenamefont {Reinhard}, \citenamefont {Almassarani}, \citenamefont {Beleites}, \citenamefont {Ronneberger}, \citenamefont {Grosse}, \citenamefont {Paulus}, \citenamefont {H{\"{u}}bner}, \citenamefont {May},\ and\ \citenamefont {Gopal}}]{Herzer2018}%
  \BibitemOpen
  \bibfield  {author} {\bibinfo {author} {\bibfnamefont {S.}~\bibnamefont {Herzer}}, \bibinfo {author} {\bibfnamefont {A.}~\bibnamefont {Woldegeorgis}}, \bibinfo {author} {\bibfnamefont {J.}~\bibnamefont {Polz}}, \bibinfo {author} {\bibfnamefont {A.}~\bibnamefont {Reinhard}}, \bibinfo {author} {\bibfnamefont {M.}~\bibnamefont {Almassarani}}, \bibinfo {author} {\bibfnamefont {B.}~\bibnamefont {Beleites}}, \bibinfo {author} {\bibfnamefont {F.}~\bibnamefont {Ronneberger}}, \bibinfo {author} {\bibfnamefont {R.}~\bibnamefont {Grosse}}, \bibinfo {author} {\bibfnamefont {G.~G.}\ \bibnamefont {Paulus}}, \bibinfo {author} {\bibfnamefont {U.}~\bibnamefont {H{\"{u}}bner}}, \bibinfo {author} {\bibfnamefont {T.}~\bibnamefont {May}},\ and\ \bibinfo {author} {\bibfnamefont {A.}~\bibnamefont {Gopal}},\ }\bibfield  {title} {\bibinfo {title} {{An investigation on THz yield from laser-produced solid density plasmas at relativistic laser intensities}},\ }\bibfield  {journal} {\bibinfo  {journal} {New Journal of Physics}\
  }\textbf {\bibinfo {volume} {20}},\ \href {https://doi.org/10.1088/1367-2630/aaada0} {10.1088/1367-2630/aaada0} (\bibinfo {year} {2018})\BibitemShut {NoStop}%
\bibitem [{\citenamefont {Khalatpour}\ \emph {et~al.}(2021)\citenamefont {Khalatpour}, \citenamefont {Paulsen}, \citenamefont {Deimert}, \citenamefont {Wasilewski},\ and\ \citenamefont {Hu}}]{Khalatpour2021High-powerSystems}%
  \BibitemOpen
  \bibfield  {author} {\bibinfo {author} {\bibfnamefont {A.}~\bibnamefont {Khalatpour}}, \bibinfo {author} {\bibfnamefont {A.~K.}\ \bibnamefont {Paulsen}}, \bibinfo {author} {\bibfnamefont {C.}~\bibnamefont {Deimert}}, \bibinfo {author} {\bibfnamefont {Z.~R.}\ \bibnamefont {Wasilewski}},\ and\ \bibinfo {author} {\bibfnamefont {Q.}~\bibnamefont {Hu}},\ }\bibfield  {title} {\bibinfo {title} {{High-power portable terahertz laser systems}},\ }\href {https://doi.org/10.1038/s41566-020-00707-5} {\bibfield  {journal} {\bibinfo  {journal} {Nature Photonics}\ }\textbf {\bibinfo {volume} {15}},\ \bibinfo {pages} {16} (\bibinfo {year} {2021})}\BibitemShut {NoStop}%
\bibitem [{\citenamefont {Liao}\ \emph {et~al.}(2020)\citenamefont {Liao}, \citenamefont {Liu}, \citenamefont {Scott}, \citenamefont {Zhang}, \citenamefont {Zhu}, \citenamefont {Zhang}, \citenamefont {Li}, \citenamefont {Armstrong}, \citenamefont {Zemaityte}, \citenamefont {Bradford}, \citenamefont {Rusby}, \citenamefont {Neely}, \citenamefont {Huggard}, \citenamefont {Mckenna}, \citenamefont {Brenner}, \citenamefont {Woolsey}, \citenamefont {Wang}, \citenamefont {Sheng},\ and\ \citenamefont {Zhang}}]{Liao2020}%
  \BibitemOpen
  \bibfield  {author} {\bibinfo {author} {\bibfnamefont {G.~Q.}\ \bibnamefont {Liao}}, \bibinfo {author} {\bibfnamefont {H.}~\bibnamefont {Liu}}, \bibinfo {author} {\bibfnamefont {G.~G.}\ \bibnamefont {Scott}}, \bibinfo {author} {\bibfnamefont {Y.~H.}\ \bibnamefont {Zhang}}, \bibinfo {author} {\bibfnamefont {B.~J.}\ \bibnamefont {Zhu}}, \bibinfo {author} {\bibfnamefont {Z.}~\bibnamefont {Zhang}}, \bibinfo {author} {\bibfnamefont {Y.~T.}\ \bibnamefont {Li}}, \bibinfo {author} {\bibfnamefont {C.}~\bibnamefont {Armstrong}}, \bibinfo {author} {\bibfnamefont {E.}~\bibnamefont {Zemaityte}}, \bibinfo {author} {\bibfnamefont {P.}~\bibnamefont {Bradford}}, \bibinfo {author} {\bibfnamefont {D.~R.}\ \bibnamefont {Rusby}}, \bibinfo {author} {\bibfnamefont {D.}~\bibnamefont {Neely}}, \bibinfo {author} {\bibfnamefont {P.~G.}\ \bibnamefont {Huggard}}, \bibinfo {author} {\bibfnamefont {P.}~\bibnamefont {Mckenna}}, \bibinfo {author} {\bibfnamefont {C.~M.}\ \bibnamefont {Brenner}}, \bibinfo {author} {\bibfnamefont {N.~C.}\
  \bibnamefont {Woolsey}}, \bibinfo {author} {\bibfnamefont {W.~M.}\ \bibnamefont {Wang}}, \bibinfo {author} {\bibfnamefont {Z.~M.}\ \bibnamefont {Sheng}},\ and\ \bibinfo {author} {\bibfnamefont {J.}~\bibnamefont {Zhang}},\ }\bibfield  {title} {\bibinfo {title} {{Towards Terawatt-Scale Spectrally Tunable Terahertz Pulses via Relativistic Laser-Foil Interactions}},\ }\href {https://doi.org/10.1103/PhysRevX.10.031062} {\bibfield  {journal} {\bibinfo  {journal} {Physical Review X}\ }\textbf {\bibinfo {volume} {10}},\ \bibinfo {pages} {31062} (\bibinfo {year} {2020})}\BibitemShut {NoStop}%
\bibitem [{\citenamefont {Lei}\ \emph {et~al.}(2022)\citenamefont {Lei}, \citenamefont {Sun}, \citenamefont {Wang}, \citenamefont {Chen}, \citenamefont {Wang}, \citenamefont {Wei}, \citenamefont {Ma}, \citenamefont {Liao},\ and\ \citenamefont {Li}}]{Lei2022HighlyInteractions}%
  \BibitemOpen
  \bibfield  {author} {\bibinfo {author} {\bibfnamefont {H.-Y.}\ \bibnamefont {Lei}}, \bibinfo {author} {\bibfnamefont {F.-Z.}\ \bibnamefont {Sun}}, \bibinfo {author} {\bibfnamefont {T.-Z.}\ \bibnamefont {Wang}}, \bibinfo {author} {\bibfnamefont {H.}~\bibnamefont {Chen}}, \bibinfo {author} {\bibfnamefont {D.}~\bibnamefont {Wang}}, \bibinfo {author} {\bibfnamefont {Y.~Y.}\ \bibnamefont {Wei}}, \bibinfo {author} {\bibfnamefont {J.~L.}\ \bibnamefont {Ma}}, \bibinfo {author} {\bibfnamefont {G.~Q.}\ \bibnamefont {Liao}},\ and\ \bibinfo {author} {\bibfnamefont {Y.~T.}\ \bibnamefont {Li}},\ }\bibfield  {title} {\bibinfo {title} {{Highly efficient generation of GV/m-level terahertz pulses from intense femtosecond laser-foil interactions}},\ }\href {https://doi.org/10.1016/j.isci.2022.104336} {\bibfield  {journal} {\bibinfo  {journal} {iScience}\ }\textbf {\bibinfo {volume} {25}},\ \bibinfo {pages} {104336} (\bibinfo {year} {2022})}\BibitemShut {NoStop}%
\bibitem [{\citenamefont {Zeng}\ \emph {et~al.}(2020)\citenamefont {Zeng}, \citenamefont {Zhou}, \citenamefont {Song}, \citenamefont {Lu}, \citenamefont {Li}, \citenamefont {Ding}, \citenamefont {Bai}, \citenamefont {Xu}, \citenamefont {Leng}, \citenamefont {Tian}, \citenamefont {Liu}, \citenamefont {Li},\ and\ \citenamefont {Xu}}]{Zeng2020}%
  \BibitemOpen
  \bibfield  {author} {\bibinfo {author} {\bibfnamefont {Y.}~\bibnamefont {Zeng}}, \bibinfo {author} {\bibfnamefont {C.}~\bibnamefont {Zhou}}, \bibinfo {author} {\bibfnamefont {L.}~\bibnamefont {Song}}, \bibinfo {author} {\bibfnamefont {X.}~\bibnamefont {Lu}}, \bibinfo {author} {\bibfnamefont {Z.}~\bibnamefont {Li}}, \bibinfo {author} {\bibfnamefont {Y.}~\bibnamefont {Ding}}, \bibinfo {author} {\bibfnamefont {Y.}~\bibnamefont {Bai}}, \bibinfo {author} {\bibfnamefont {Y.}~\bibnamefont {Xu}}, \bibinfo {author} {\bibfnamefont {Y.}~\bibnamefont {Leng}}, \bibinfo {author} {\bibfnamefont {Y.}~\bibnamefont {Tian}}, \bibinfo {author} {\bibfnamefont {J.}~\bibnamefont {Liu}}, \bibinfo {author} {\bibfnamefont {R.}~\bibnamefont {Li}},\ and\ \bibinfo {author} {\bibfnamefont {Z.}~\bibnamefont {Xu}},\ }\bibfield  {title} {\bibinfo {title} {{Guiding and emission of milijoule single-cycle THz pulse from laser-driven wire-like targets}},\ }\href {https://doi.org/10.1364/oe.390764} {\bibfield  {journal} {\bibinfo  {journal}
  {Optics Express}\ }\textbf {\bibinfo {volume} {28}},\ \bibinfo {pages} {15258} (\bibinfo {year} {2020})}\BibitemShut {NoStop}%
\bibitem [{\citenamefont {F{\"{u}}l{\"{o}}p}\ \emph {et~al.}(2020)\citenamefont {F{\"{u}}l{\"{o}}p}, \citenamefont {Tzortzakis},\ and\ \citenamefont {Kampfrath}}]{Fulop2020Laser-DrivenSources}%
  \BibitemOpen
  \bibfield  {author} {\bibinfo {author} {\bibfnamefont {J.~A.}\ \bibnamefont {F{\"{u}}l{\"{o}}p}}, \bibinfo {author} {\bibfnamefont {S.}~\bibnamefont {Tzortzakis}},\ and\ \bibinfo {author} {\bibfnamefont {T.}~\bibnamefont {Kampfrath}},\ }\bibfield  {title} {\bibinfo {title} {{Laser-Driven Strong-Field Terahertz Sources}},\ }\href {https://doi.org/10.1002/adom.201900681} {\bibfield  {journal} {\bibinfo  {journal} {Advanced Optical Materials}\ }\textbf {\bibinfo {volume} {8}},\ \bibinfo {pages} {1} (\bibinfo {year} {2020})}\BibitemShut {NoStop}%
\bibitem [{\citenamefont {Sal{\'{e}}n}\ \emph {et~al.}(2019)\citenamefont {Sal{\'{e}}n}, \citenamefont {Basini}, \citenamefont {Bonetti}, \citenamefont {Hebling}, \citenamefont {Krasilnikov}, \citenamefont {Nikitin}, \citenamefont {Shamuilov}, \citenamefont {Tibai}, \citenamefont {Zhaunerchyk},\ and\ \citenamefont {Goryashko}}]{Salen2019MatterTechnology}%
  \BibitemOpen
  \bibfield  {author} {\bibinfo {author} {\bibfnamefont {P.}~\bibnamefont {Sal{\'{e}}n}}, \bibinfo {author} {\bibfnamefont {M.}~\bibnamefont {Basini}}, \bibinfo {author} {\bibfnamefont {S.}~\bibnamefont {Bonetti}}, \bibinfo {author} {\bibfnamefont {J.}~\bibnamefont {Hebling}}, \bibinfo {author} {\bibfnamefont {M.}~\bibnamefont {Krasilnikov}}, \bibinfo {author} {\bibfnamefont {A.~Y.}\ \bibnamefont {Nikitin}}, \bibinfo {author} {\bibfnamefont {G.}~\bibnamefont {Shamuilov}}, \bibinfo {author} {\bibfnamefont {Z.}~\bibnamefont {Tibai}}, \bibinfo {author} {\bibfnamefont {V.}~\bibnamefont {Zhaunerchyk}},\ and\ \bibinfo {author} {\bibfnamefont {V.}~\bibnamefont {Goryashko}},\ }\bibfield  {title} {\bibinfo {title} {{Matter manipulation with extreme terahertz light: Progress in the enabling THz technology}},\ }\href {https://doi.org/10.1016/j.physrep.2019.09.002} {\bibfield  {journal} {\bibinfo  {journal} {Physics Reports}\ }\textbf {\bibinfo {volume} {836-837}},\ \bibinfo {pages} {1} (\bibinfo {year}
  {2019})}\BibitemShut {NoStop}%
\bibitem [{\citenamefont {Hamm}\ \emph {et~al.}(2017)\citenamefont {Hamm}, \citenamefont {Meuwly}, \citenamefont {Johnson}, \citenamefont {Beaud},\ and\ \citenamefont {Staub}}]{Hamm2017Perspective:Molecules}%
  \BibitemOpen
  \bibfield  {author} {\bibinfo {author} {\bibfnamefont {P.}~\bibnamefont {Hamm}}, \bibinfo {author} {\bibfnamefont {M.}~\bibnamefont {Meuwly}}, \bibinfo {author} {\bibfnamefont {S.~L.}\ \bibnamefont {Johnson}}, \bibinfo {author} {\bibfnamefont {P.}~\bibnamefont {Beaud}},\ and\ \bibinfo {author} {\bibfnamefont {U.}~\bibnamefont {Staub}},\ }\bibfield  {title} {\bibinfo {title} {{Perspective: THz-driven nuclear dynamics from solids to molecules}},\ }\bibfield  {journal} {\bibinfo  {journal} {Structural Dynamics}\ }\textbf {\bibinfo {volume} {4}},\ \href {https://doi.org/10.1063/1.4992050} {10.1063/1.4992050} (\bibinfo {year} {2017})\BibitemShut {NoStop}%
\bibitem [{\citenamefont {Zhang}\ \emph {et~al.}(2021{\natexlab{a}})\citenamefont {Zhang}, \citenamefont {Li},\ and\ \citenamefont {Zhao}}]{Zhang2021IntenseApplication}%
  \BibitemOpen
  \bibfield  {author} {\bibinfo {author} {\bibfnamefont {Y.}~\bibnamefont {Zhang}}, \bibinfo {author} {\bibfnamefont {K.}~\bibnamefont {Li}},\ and\ \bibinfo {author} {\bibfnamefont {H.}~\bibnamefont {Zhao}},\ }\bibfield  {title} {\bibinfo {title} {{Intense terahertz radiation: generation and application}},\ }\href {https://doi.org/10.1007/s12200-020-1052-9} {\bibfield  {journal} {\bibinfo  {journal} {Frontiers of Optoelectronics}\ }\textbf {\bibinfo {volume} {14}},\ \bibinfo {pages} {4} (\bibinfo {year} {2021}{\natexlab{a}})}\BibitemShut {NoStop}%
\bibitem [{\citenamefont {Zhang}\ \emph {et~al.}(2017)\citenamefont {Zhang}, \citenamefont {Shkurinov},\ and\ \citenamefont {Zhang}}]{Zhang2017ExtremeScience}%
  \BibitemOpen
  \bibfield  {author} {\bibinfo {author} {\bibfnamefont {X.~C.}\ \bibnamefont {Zhang}}, \bibinfo {author} {\bibfnamefont {A.}~\bibnamefont {Shkurinov}},\ and\ \bibinfo {author} {\bibfnamefont {Y.}~\bibnamefont {Zhang}},\ }\bibfield  {title} {\bibinfo {title} {{Extreme terahertz science}},\ }\href {https://doi.org/10.1038/nphoton.2016.249} {\bibfield  {journal} {\bibinfo  {journal} {Nature Photonics}\ }\textbf {\bibinfo {volume} {11}},\ \bibinfo {pages} {16} (\bibinfo {year} {2017})}\BibitemShut {NoStop}%
\bibitem [{\citenamefont {Zhang}\ \emph {et~al.}(2021{\natexlab{b}})\citenamefont {Zhang}, \citenamefont {Ma}, \citenamefont {Ma}, \citenamefont {Wu}, \citenamefont {Ouyang}, \citenamefont {Kong}, \citenamefont {Hong}, \citenamefont {Wang}, \citenamefont {Yang}, \citenamefont {Chen}, \citenamefont {Li},\ and\ \citenamefont {Zhang}}]{Zhang20211.4-mJNiobates}%
  \BibitemOpen
  \bibfield  {author} {\bibinfo {author} {\bibfnamefont {B.}~\bibnamefont {Zhang}}, \bibinfo {author} {\bibfnamefont {Z.}~\bibnamefont {Ma}}, \bibinfo {author} {\bibfnamefont {J.}~\bibnamefont {Ma}}, \bibinfo {author} {\bibfnamefont {X.}~\bibnamefont {Wu}}, \bibinfo {author} {\bibfnamefont {C.}~\bibnamefont {Ouyang}}, \bibinfo {author} {\bibfnamefont {D.}~\bibnamefont {Kong}}, \bibinfo {author} {\bibfnamefont {T.}~\bibnamefont {Hong}}, \bibinfo {author} {\bibfnamefont {X.}~\bibnamefont {Wang}}, \bibinfo {author} {\bibfnamefont {P.}~\bibnamefont {Yang}}, \bibinfo {author} {\bibfnamefont {L.}~\bibnamefont {Chen}}, \bibinfo {author} {\bibfnamefont {Y.}~\bibnamefont {Li}},\ and\ \bibinfo {author} {\bibfnamefont {J.}~\bibnamefont {Zhang}},\ }\bibfield  {title} {\bibinfo {title} {{1.4-mJ High Energy Terahertz Radiation from Lithium Niobates}},\ }\href {https://doi.org/10.1002/lpor.202000295} {\bibfield  {journal} {\bibinfo  {journal} {Laser and Photonics Reviews}\ }\textbf {\bibinfo {volume} {15}},\ \bibinfo
  {pages} {1} (\bibinfo {year} {2021}{\natexlab{b}})}\BibitemShut {NoStop}%
\bibitem [{\citenamefont {D{\'{e}}chard}\ \emph {et~al.}(2018)\citenamefont {D{\'{e}}chard}, \citenamefont {Debayle}, \citenamefont {Davoine}, \citenamefont {Gremillet},\ and\ \citenamefont {Berg{\'{e}}}}]{Dechard2018TerahertzRadiation}%
  \BibitemOpen
  \bibfield  {author} {\bibinfo {author} {\bibfnamefont {J.}~\bibnamefont {D{\'{e}}chard}}, \bibinfo {author} {\bibfnamefont {A.}~\bibnamefont {Debayle}}, \bibinfo {author} {\bibfnamefont {X.}~\bibnamefont {Davoine}}, \bibinfo {author} {\bibfnamefont {L.}~\bibnamefont {Gremillet}},\ and\ \bibinfo {author} {\bibfnamefont {L.}~\bibnamefont {Berg{\'{e}}}},\ }\bibfield  {title} {\bibinfo {title} {{Terahertz Pulse Generation in Underdense Relativistic Plasmas: From Photoionization-Induced Radiation to Coherent Transition Radiation}},\ }\href {https://doi.org/10.1103/PhysRevLett.120.144801} {\bibfield  {journal} {\bibinfo  {journal} {Physical Review Letters}\ }\textbf {\bibinfo {volume} {120}},\ \bibinfo {pages} {6} (\bibinfo {year} {2018})}\BibitemShut {NoStop}%
\bibitem [{\citenamefont {E}\ \emph {et~al.}(2021)\citenamefont {E}, \citenamefont {Zhang}, \citenamefont {Tcypkin}, \citenamefont {Kozlov}, \citenamefont {Zhang},\ and\ \citenamefont {Zhang}}]{E2021BroadbandLiquids}%
  \BibitemOpen
  \bibfield  {author} {\bibinfo {author} {\bibfnamefont {Y.}~\bibnamefont {E}}, \bibinfo {author} {\bibfnamefont {L.}~\bibnamefont {Zhang}}, \bibinfo {author} {\bibfnamefont {A.}~\bibnamefont {Tcypkin}}, \bibinfo {author} {\bibfnamefont {S.}~\bibnamefont {Kozlov}}, \bibinfo {author} {\bibfnamefont {C.}~\bibnamefont {Zhang}},\ and\ \bibinfo {author} {\bibfnamefont {X.~C.}\ \bibnamefont {Zhang}},\ }\bibfield  {title} {\bibinfo {title} {{Broadband THz Sources from Gases to Liquids}},\ }\bibfield  {journal} {\bibinfo  {journal} {Ultrafast Science}\ }\textbf {\bibinfo {volume} {2021}},\ \href {https://doi.org/10.34133/2021/9892763} {10.34133/2021/9892763} (\bibinfo {year} {2021})\BibitemShut {NoStop}%
\bibitem [{\citenamefont {Zhang}\ \emph {et~al.}(2022)\citenamefont {Zhang}, \citenamefont {Zeng}, \citenamefont {Bai}, \citenamefont {Li}, \citenamefont {Tian},\ and\ \citenamefont {Li}}]{Zhang2022}%
  \BibitemOpen
  \bibfield  {author} {\bibinfo {author} {\bibfnamefont {D.}~\bibnamefont {Zhang}}, \bibinfo {author} {\bibfnamefont {Y.}~\bibnamefont {Zeng}}, \bibinfo {author} {\bibfnamefont {Y.}~\bibnamefont {Bai}}, \bibinfo {author} {\bibfnamefont {Z.}~\bibnamefont {Li}}, \bibinfo {author} {\bibfnamefont {Y.}~\bibnamefont {Tian}},\ and\ \bibinfo {author} {\bibfnamefont {R.}~\bibnamefont {Li}},\ }\bibfield  {title} {\bibinfo {title} {{Coherent surface plasmon polariton amplification via free-electron pumping}},\ }\href {https://doi.org/10.1038/s41586-022-05239-2} {\bibfield  {journal} {\bibinfo  {journal} {Nature}\ }\textbf {\bibinfo {volume} {611}},\ \bibinfo {pages} {55} (\bibinfo {year} {2022})}\BibitemShut {NoStop}%
\bibitem [{\citenamefont {Zhuo}\ \emph {et~al.}(2017)\citenamefont {Zhuo}, \citenamefont {Zhang}, \citenamefont {Li}, \citenamefont {Zhou}, \citenamefont {Li}, \citenamefont {Zou}, \citenamefont {Yu}, \citenamefont {Wu}, \citenamefont {Sheng},\ and\ \citenamefont {Zhou}}]{Zhuo2017}%
  \BibitemOpen
  \bibfield  {author} {\bibinfo {author} {\bibfnamefont {H.~B.}\ \bibnamefont {Zhuo}}, \bibinfo {author} {\bibfnamefont {S.~J.}\ \bibnamefont {Zhang}}, \bibinfo {author} {\bibfnamefont {X.~H.}\ \bibnamefont {Li}}, \bibinfo {author} {\bibfnamefont {H.~Y.}\ \bibnamefont {Zhou}}, \bibinfo {author} {\bibfnamefont {X.~Z.}\ \bibnamefont {Li}}, \bibinfo {author} {\bibfnamefont {D.~B.}\ \bibnamefont {Zou}}, \bibinfo {author} {\bibfnamefont {M.~Y.}\ \bibnamefont {Yu}}, \bibinfo {author} {\bibfnamefont {H.~C.}\ \bibnamefont {Wu}}, \bibinfo {author} {\bibfnamefont {Z.~M.}\ \bibnamefont {Sheng}},\ and\ \bibinfo {author} {\bibfnamefont {C.~T.}\ \bibnamefont {Zhou}},\ }\bibfield  {title} {\bibinfo {title} {{Terahertz generation from laser-driven ultrafast current propagation along a wire target}},\ }\href {https://doi.org/10.1103/PhysRevE.95.013201} {\bibfield  {journal} {\bibinfo  {journal} {Physical Review E}\ }\textbf {\bibinfo {volume} {95}},\ \bibinfo {pages} {2} (\bibinfo {year} {2017})}\BibitemShut {NoStop}%
\bibitem [{\citenamefont {Kumar}\ \emph {et~al.}(2022)\citenamefont {Kumar}, \citenamefont {Gupta}, \citenamefont {Srivastava}, \citenamefont {Devi}, \citenamefont {Kumar},\ and\ \citenamefont {Roy~Chowdhury}}]{Kumar2022PhotoinducedParameters}%
  \BibitemOpen
  \bibfield  {author} {\bibinfo {author} {\bibfnamefont {D.}~\bibnamefont {Kumar}}, \bibinfo {author} {\bibfnamefont {M.}~\bibnamefont {Gupta}}, \bibinfo {author} {\bibfnamefont {Y.~K.}\ \bibnamefont {Srivastava}}, \bibinfo {author} {\bibfnamefont {K.~M.}\ \bibnamefont {Devi}}, \bibinfo {author} {\bibfnamefont {R.}~\bibnamefont {Kumar}},\ and\ \bibinfo {author} {\bibfnamefont {D.}~\bibnamefont {Roy~Chowdhury}},\ }\bibfield  {title} {\bibinfo {title} {{Photoinduced dynamic tailoring of near-field coupled terahertz metasurfaces and its effect on Coulomb parameters}},\ }\bibfield  {journal} {\bibinfo  {journal} {Journal of Optics (United Kingdom)}\ }\textbf {\bibinfo {volume} {24}},\ \href {https://doi.org/10.1088/2040-8986/ac4d71} {10.1088/2040-8986/ac4d71} (\bibinfo {year} {2022})\BibitemShut {NoStop}%
\bibitem [{\citenamefont {Kang}\ \emph {et~al.}(2024)\citenamefont {Kang}, \citenamefont {Lee}, \citenamefont {Kim}, \citenamefont {Yang}, \citenamefont {Nam}, \citenamefont {Kim}, \citenamefont {Baek}, \citenamefont {Yoon}, \citenamefont {Lee}, \citenamefont {Kim},\ and\ \citenamefont {Kim}}]{Kang2024FrequencyMetamaterials}%
  \BibitemOpen
  \bibfield  {author} {\bibinfo {author} {\bibfnamefont {G.}~\bibnamefont {Kang}}, \bibinfo {author} {\bibfnamefont {Y.}~\bibnamefont {Lee}}, \bibinfo {author} {\bibfnamefont {J.}~\bibnamefont {Kim}}, \bibinfo {author} {\bibfnamefont {D.}~\bibnamefont {Yang}}, \bibinfo {author} {\bibfnamefont {H.~K.}\ \bibnamefont {Nam}}, \bibinfo {author} {\bibfnamefont {S.}~\bibnamefont {Kim}}, \bibinfo {author} {\bibfnamefont {S.}~\bibnamefont {Baek}}, \bibinfo {author} {\bibfnamefont {H.}~\bibnamefont {Yoon}}, \bibinfo {author} {\bibfnamefont {J.}~\bibnamefont {Lee}}, \bibinfo {author} {\bibfnamefont {T.~T.}\ \bibnamefont {Kim}},\ and\ \bibinfo {author} {\bibfnamefont {Y.~J.}\ \bibnamefont {Kim}},\ }\bibfield  {title} {\bibinfo {title} {{Frequency comb measurements for 6G terahertz nano/microphotonics and metamaterials}},\ }\href {https://doi.org/10.1515/nanoph-2023-0869} {\bibfield  {journal} {\bibinfo  {journal} {Nanophotonics}\ ,\ \bibinfo {pages} {1}} (\bibinfo {year} {2024})}\BibitemShut {NoStop}%
\bibitem [{\citenamefont {Rinderknecht}\ \emph {et~al.}(2021)\citenamefont {Rinderknecht}, \citenamefont {Wang}, \citenamefont {Garcia}, \citenamefont {Bruhaug}, \citenamefont {Wei}, \citenamefont {Quevedo}, \citenamefont {Ditmire}, \citenamefont {Williams}, \citenamefont {Haid}, \citenamefont {Doria}, \citenamefont {Spohr}, \citenamefont {Toncian},\ and\ \citenamefont {Arefiev}}]{Rinderknecht2021}%
  \BibitemOpen
  \bibfield  {author} {\bibinfo {author} {\bibfnamefont {H.~G.}\ \bibnamefont {Rinderknecht}}, \bibinfo {author} {\bibfnamefont {T.}~\bibnamefont {Wang}}, \bibinfo {author} {\bibfnamefont {A.~L.}\ \bibnamefont {Garcia}}, \bibinfo {author} {\bibfnamefont {G.}~\bibnamefont {Bruhaug}}, \bibinfo {author} {\bibfnamefont {M.~S.}\ \bibnamefont {Wei}}, \bibinfo {author} {\bibfnamefont {H.~J.}\ \bibnamefont {Quevedo}}, \bibinfo {author} {\bibfnamefont {T.}~\bibnamefont {Ditmire}}, \bibinfo {author} {\bibfnamefont {J.}~\bibnamefont {Williams}}, \bibinfo {author} {\bibfnamefont {A.}~\bibnamefont {Haid}}, \bibinfo {author} {\bibfnamefont {D.}~\bibnamefont {Doria}}, \bibinfo {author} {\bibfnamefont {K.~M.}\ \bibnamefont {Spohr}}, \bibinfo {author} {\bibfnamefont {T.}~\bibnamefont {Toncian}},\ and\ \bibinfo {author} {\bibfnamefont {A.}~\bibnamefont {Arefiev}},\ }\bibfield  {title} {\bibinfo {title} {{Relativistically transparent magnetic filaments: Scaling laws, initial results and prospects for strong-field QED
  studies}},\ }\bibfield  {journal} {\bibinfo  {journal} {New Journal of Physics}\ }\textbf {\bibinfo {volume} {23}},\ \href {https://doi.org/10.1088/1367-2630/ac22e7} {10.1088/1367-2630/ac22e7} (\bibinfo {year} {2021})\BibitemShut {NoStop}%
\bibitem [{\citenamefont {Yi}\ and\ \citenamefont {F{\"{u}}l{\"{o}}p}(2019)}]{Yi2019}%
  \BibitemOpen
  \bibfield  {author} {\bibinfo {author} {\bibfnamefont {L.}~\bibnamefont {Yi}}\ and\ \bibinfo {author} {\bibfnamefont {T.}~\bibnamefont {F{\"{u}}l{\"{o}}p}},\ }\bibfield  {title} {\bibinfo {title} {{Coherent Diffraction Radiation of Relativistic Terahertz Pulses from a Laser-Driven Microplasma Waveguide}},\ }\href {https://doi.org/10.1103/PhysRevLett.123.094801} {\bibfield  {journal} {\bibinfo  {journal} {Physical Review Letters}\ }\textbf {\bibinfo {volume} {123}},\ \bibinfo {pages} {94801} (\bibinfo {year} {2019})}\BibitemShut {NoStop}%
\bibitem [{\citenamefont {Rosmej}\ \emph {et~al.}(2019)\citenamefont {Rosmej}, \citenamefont {Andreev}, \citenamefont {Zaehter}, \citenamefont {Zahn}, \citenamefont {Christ}, \citenamefont {Borm}, \citenamefont {Radon}, \citenamefont {Sokolov}, \citenamefont {Pugachev}, \citenamefont {Khaghani}, \citenamefont {Horst}, \citenamefont {Borisenko}, \citenamefont {Sklizkov},\ and\ \citenamefont {Pimenov}}]{Rosmej2019InteractionGamma-rays}%
  \BibitemOpen
  \bibfield  {author} {\bibinfo {author} {\bibfnamefont {O.~N.}\ \bibnamefont {Rosmej}}, \bibinfo {author} {\bibfnamefont {N.~E.}\ \bibnamefont {Andreev}}, \bibinfo {author} {\bibfnamefont {S.}~\bibnamefont {Zaehter}}, \bibinfo {author} {\bibfnamefont {N.}~\bibnamefont {Zahn}}, \bibinfo {author} {\bibfnamefont {P.}~\bibnamefont {Christ}}, \bibinfo {author} {\bibfnamefont {B.}~\bibnamefont {Borm}}, \bibinfo {author} {\bibfnamefont {T.}~\bibnamefont {Radon}}, \bibinfo {author} {\bibfnamefont {A.}~\bibnamefont {Sokolov}}, \bibinfo {author} {\bibfnamefont {L.~P.}\ \bibnamefont {Pugachev}}, \bibinfo {author} {\bibfnamefont {D.}~\bibnamefont {Khaghani}}, \bibinfo {author} {\bibfnamefont {F.}~\bibnamefont {Horst}}, \bibinfo {author} {\bibfnamefont {N.~G.}\ \bibnamefont {Borisenko}}, \bibinfo {author} {\bibfnamefont {G.}~\bibnamefont {Sklizkov}},\ and\ \bibinfo {author} {\bibfnamefont {V.~G.}\ \bibnamefont {Pimenov}},\ }\bibfield  {title} {\bibinfo {title} {{Interaction of relativistically intense laser pulses with
  long-scale near critical plasmas for optimization of laser based sources of MeV electrons and gamma-rays}},\ }\bibfield  {journal} {\bibinfo  {journal} {New Journal of Physics}\ }\textbf {\bibinfo {volume} {21}},\ \href {https://doi.org/10.1088/1367-2630/ab1047} {10.1088/1367-2630/ab1047} (\bibinfo {year} {2019})\BibitemShut {NoStop}%
\bibitem [{\citenamefont {Bruhaug}\ \emph {et~al.}(2022)\citenamefont {Bruhaug}, \citenamefont {Rinderknecht}, \citenamefont {E}, \citenamefont {Brannon}, \citenamefont {Guy}, \citenamefont {Peck}, \citenamefont {Landis}, \citenamefont {Brent}, \citenamefont {Fairbanks}, \citenamefont {Mcatee}, \citenamefont {Walker}, \citenamefont {Buczek}, \citenamefont {Krieger}, \citenamefont {Romanofsky}, \citenamefont {Mileham}, \citenamefont {Francis}, \citenamefont {Zhang}, \citenamefont {Collins},\ and\ \citenamefont {Rygg}}]{Bruhaug2022DevelopmentLaser}%
  \BibitemOpen
  \bibfield  {author} {\bibinfo {author} {\bibfnamefont {G.}~\bibnamefont {Bruhaug}}, \bibinfo {author} {\bibfnamefont {H.~G.}\ \bibnamefont {Rinderknecht}}, \bibinfo {author} {\bibfnamefont {Y.}~\bibnamefont {E}}, \bibinfo {author} {\bibfnamefont {R.~B.}\ \bibnamefont {Brannon}}, \bibinfo {author} {\bibfnamefont {D.}~\bibnamefont {Guy}}, \bibinfo {author} {\bibfnamefont {R.~G.}\ \bibnamefont {Peck}}, \bibinfo {author} {\bibfnamefont {N.}~\bibnamefont {Landis}}, \bibinfo {author} {\bibfnamefont {G.}~\bibnamefont {Brent}}, \bibinfo {author} {\bibfnamefont {R.}~\bibnamefont {Fairbanks}}, \bibinfo {author} {\bibfnamefont {C.}~\bibnamefont {Mcatee}}, \bibinfo {author} {\bibfnamefont {T.}~\bibnamefont {Walker}}, \bibinfo {author} {\bibfnamefont {T.}~\bibnamefont {Buczek}}, \bibinfo {author} {\bibfnamefont {M.}~\bibnamefont {Krieger}}, \bibinfo {author} {\bibfnamefont {M.~H.}\ \bibnamefont {Romanofsky}}, \bibinfo {author} {\bibfnamefont {C.}~\bibnamefont {Mileham}}, \bibinfo {author} {\bibfnamefont {K.~G.}\
  \bibnamefont {Francis}}, \bibinfo {author} {\bibfnamefont {X.-c.}\ \bibnamefont {Zhang}}, \bibinfo {author} {\bibfnamefont {G.~W.}\ \bibnamefont {Collins}},\ and\ \bibinfo {author} {\bibfnamefont {J.~R.}\ \bibnamefont {Rygg}},\ }\bibfield  {title} {\bibinfo {title} {{Development of a hardened THz energy meter for use on the kilojoule-scale , short- pulse OMEGA EP laser Development of a hardened THz energy meter for use on the kilojoule-scale , short-pulse OMEGA EP laser}},\ }\href {https://doi.org/10.1063/5.0099328} {\bibfield  {journal} {\bibinfo  {journal} {Review of Scientific Instruments}\ }\textbf {\bibinfo {volume} {123502}},\ \bibinfo {pages} {93} (\bibinfo {year} {2022})}\BibitemShut {NoStop}%
\bibitem [{\citenamefont {Ayers}(2010)}]{Ayers2010}%
  \BibitemOpen
  \bibfield  {author} {\bibinfo {author} {\bibfnamefont {S.~L.}\ \bibnamefont {Ayers}},\ }\bibfield  {title} {\bibinfo {title} {{Electron Positron Proton Spectrometer for use at Laboratory for Laser Energetics}},\ }\href@noop {} {\bibfield  {journal} {\bibinfo  {journal} {LLNL-TR-427769}\ } (\bibinfo {year} {2010})}\BibitemShut {NoStop}%
\bibitem [{\citenamefont {Nilson}\ \emph {et~al.}(2011)\citenamefont {Nilson}, \citenamefont {Solodov}, \citenamefont {Myatt}, \citenamefont {Theobald}, \citenamefont {Jaanimagi}, \citenamefont {Gao}, \citenamefont {Stoeckl}, \citenamefont {Craxton}, \citenamefont {Delettrez}, \citenamefont {Yaakobi}, \citenamefont {Zuegel}, \citenamefont {Kruschwitz}, \citenamefont {Dorrer}, \citenamefont {Kelly}, \citenamefont {Akli}, \citenamefont {Patel}, \citenamefont {MacKinnon}, \citenamefont {Betti}, \citenamefont {Sangster},\ and\ \citenamefont {Meyerhofer}}]{Nilson2011ScalingInteractions}%
  \BibitemOpen
  \bibfield  {author} {\bibinfo {author} {\bibfnamefont {P.~M.}\ \bibnamefont {Nilson}}, \bibinfo {author} {\bibfnamefont {A.~A.}\ \bibnamefont {Solodov}}, \bibinfo {author} {\bibfnamefont {J.~F.}\ \bibnamefont {Myatt}}, \bibinfo {author} {\bibfnamefont {W.}~\bibnamefont {Theobald}}, \bibinfo {author} {\bibfnamefont {P.~A.}\ \bibnamefont {Jaanimagi}}, \bibinfo {author} {\bibfnamefont {L.}~\bibnamefont {Gao}}, \bibinfo {author} {\bibfnamefont {C.}~\bibnamefont {Stoeckl}}, \bibinfo {author} {\bibfnamefont {R.~S.}\ \bibnamefont {Craxton}}, \bibinfo {author} {\bibfnamefont {J.~A.}\ \bibnamefont {Delettrez}}, \bibinfo {author} {\bibfnamefont {B.}~\bibnamefont {Yaakobi}}, \bibinfo {author} {\bibfnamefont {J.~D.}\ \bibnamefont {Zuegel}}, \bibinfo {author} {\bibfnamefont {B.~E.}\ \bibnamefont {Kruschwitz}}, \bibinfo {author} {\bibfnamefont {C.}~\bibnamefont {Dorrer}}, \bibinfo {author} {\bibfnamefont {J.~H.}\ \bibnamefont {Kelly}}, \bibinfo {author} {\bibfnamefont {K.~U.}\ \bibnamefont {Akli}}, \bibinfo {author}
  {\bibfnamefont {P.~K.}\ \bibnamefont {Patel}}, \bibinfo {author} {\bibfnamefont {A.~J.}\ \bibnamefont {MacKinnon}}, \bibinfo {author} {\bibfnamefont {R.}~\bibnamefont {Betti}}, \bibinfo {author} {\bibfnamefont {T.~C.}\ \bibnamefont {Sangster}},\ and\ \bibinfo {author} {\bibfnamefont {D.~D.}\ \bibnamefont {Meyerhofer}},\ }\bibfield  {title} {\bibinfo {title} {{Scaling hot-electron generation to long-pulse, high-intensity lasersolid interactions}},\ }\bibfield  {journal} {\bibinfo  {journal} {Physics of Plasmas}\ }\textbf {\bibinfo {volume} {18}},\ \href {https://doi.org/10.1063/1.3560569} {10.1063/1.3560569} (\bibinfo {year} {2011})\BibitemShut {NoStop}%
\bibitem [{\citenamefont {Myatt}\ \emph {et~al.}(2007)\citenamefont {Myatt}, \citenamefont {Theobald}, \citenamefont {Delettrez}, \citenamefont {Stoeckl}, \citenamefont {Storm}, \citenamefont {Sangster}, \citenamefont {Maximov},\ and\ \citenamefont {Short}}]{Myatt2007High-intensityEP}%
  \BibitemOpen
  \bibfield  {author} {\bibinfo {author} {\bibfnamefont {J.}~\bibnamefont {Myatt}}, \bibinfo {author} {\bibfnamefont {W.}~\bibnamefont {Theobald}}, \bibinfo {author} {\bibfnamefont {J.~A.}\ \bibnamefont {Delettrez}}, \bibinfo {author} {\bibfnamefont {C.}~\bibnamefont {Stoeckl}}, \bibinfo {author} {\bibfnamefont {M.}~\bibnamefont {Storm}}, \bibinfo {author} {\bibfnamefont {T.~C.}\ \bibnamefont {Sangster}}, \bibinfo {author} {\bibfnamefont {A.~V.}\ \bibnamefont {Maximov}},\ and\ \bibinfo {author} {\bibfnamefont {R.~W.}\ \bibnamefont {Short}},\ }\bibfield  {title} {\bibinfo {title} {{High-intensity laser interactions with mass-limited solid targets and implications for fast-ignition experiments on OMEGA EP}},\ }\bibfield  {journal} {\bibinfo  {journal} {Physics of Plasmas}\ }\textbf {\bibinfo {volume} {14}},\ \href {https://doi.org/10.1063/1.2472371} {10.1063/1.2472371} (\bibinfo {year} {2007})\BibitemShut {NoStop}%
\bibitem [{\citenamefont {Roth}\ and\ \citenamefont {Schollmeier}(2016)}]{Roth2016}%
  \BibitemOpen
  \bibfield  {author} {\bibinfo {author} {\bibfnamefont {M.}~\bibnamefont {Roth}}\ and\ \bibinfo {author} {\bibfnamefont {M.}~\bibnamefont {Schollmeier}},\ }\bibfield  {title} {\bibinfo {title} {{Ion Acceleration - Target Normal Sheath Acceleration (CERN Yellow Reports)}},\ }\href@noop {} {\bibfield  {journal} {\bibinfo  {journal} {Proceedings of the CAS-CERN Accelerator School: Plasma Wake Acceleration, Geneva, Switzerland}\ }\textbf {\bibinfo {volume} {001}},\ \bibinfo {pages} {Volume 1 231 } (\bibinfo {year} {2016})}\BibitemShut {NoStop}%
\bibitem [{\citenamefont {Denoual}\ \emph {et~al.}(2023)\citenamefont {Denoual}, \citenamefont {Berg{\'{e}}}, \citenamefont {Davoine},\ and\ \citenamefont {Gremillet}}]{Denoual2023ModelingPulses}%
  \BibitemOpen
  \bibfield  {author} {\bibinfo {author} {\bibfnamefont {E.}~\bibnamefont {Denoual}}, \bibinfo {author} {\bibfnamefont {L.}~\bibnamefont {Berg{\'{e}}}}, \bibinfo {author} {\bibfnamefont {X.}~\bibnamefont {Davoine}},\ and\ \bibinfo {author} {\bibfnamefont {L.}~\bibnamefont {Gremillet}},\ }\bibfield  {title} {\bibinfo {title} {{Modeling terahertz emissions from energetic electrons and ions in foil targets irradiated by ultraintense femtosecond laser pulses}},\ }\href {https://doi.org/10.1103/physreve.108.065211} {\bibfield  {journal} {\bibinfo  {journal} {Physical Review E}\ }\textbf {\bibinfo {volume} {108}},\ \bibinfo {pages} {1} (\bibinfo {year} {2023})}\BibitemShut {NoStop}%
\bibitem [{\citenamefont {Liao}\ and\ \citenamefont {Li}(2023)}]{Liao2023PerspectivesPlasmas}%
  \BibitemOpen
  \bibfield  {author} {\bibinfo {author} {\bibfnamefont {G.}~\bibnamefont {Liao}}\ and\ \bibinfo {author} {\bibfnamefont {Y.}~\bibnamefont {Li}},\ }\bibfield  {title} {\bibinfo {title} {{Perspectives on ultraintense laser-driven terahertz radiation from plasmas}},\ }\bibfield  {journal} {\bibinfo  {journal} {Physics of Plasmas}\ }\textbf {\bibinfo {volume} {30}},\ \href {https://doi.org/10.1063/5.0167730} {10.1063/5.0167730} (\bibinfo {year} {2023})\BibitemShut {NoStop}%
\bibitem [{\citenamefont {Schroeder}\ \emph {et~al.}(2004)\citenamefont {Schroeder}, \citenamefont {Esarey}, \citenamefont {van Tilborg},\ and\ \citenamefont {Leemans}}]{Schroeder2004}%
  \BibitemOpen
  \bibfield  {author} {\bibinfo {author} {\bibfnamefont {C.~B.}\ \bibnamefont {Schroeder}}, \bibinfo {author} {\bibfnamefont {E.}~\bibnamefont {Esarey}}, \bibinfo {author} {\bibfnamefont {J.}~\bibnamefont {van Tilborg}},\ and\ \bibinfo {author} {\bibfnamefont {W.~P.}\ \bibnamefont {Leemans}},\ }\bibfield  {title} {\bibinfo {title} {{Theory of coherent transition radiation generated at a plasma-vacuum interface}},\ }\href {https://doi.org/10.1103/PhysRevE.69.016501} {\bibfield  {journal} {\bibinfo  {journal} {Physical Review E - Statistical Physics, Plasmas, Fluids, and Related Interdisciplinary Topics}\ }\textbf {\bibinfo {volume} {69}},\ \bibinfo {pages} {12} (\bibinfo {year} {2004})}\BibitemShut {NoStop}%
\bibitem [{\citenamefont {Ping}\ \emph {et~al.}(2008)\citenamefont {Ping}, \citenamefont {Shepherd}, \citenamefont {Lasinski}, \citenamefont {Tabak}, \citenamefont {Chen}, \citenamefont {Chung}, \citenamefont {Fournier}, \citenamefont {Hansen}, \citenamefont {Kemp}, \citenamefont {Liedahl}, \citenamefont {Widmann}, \citenamefont {Wilks}, \citenamefont {Rozmus},\ and\ \citenamefont {Sherlock}}]{Ping2008AbsorptionRegime}%
  \BibitemOpen
  \bibfield  {author} {\bibinfo {author} {\bibfnamefont {Y.}~\bibnamefont {Ping}}, \bibinfo {author} {\bibfnamefont {R.}~\bibnamefont {Shepherd}}, \bibinfo {author} {\bibfnamefont {B.~F.}\ \bibnamefont {Lasinski}}, \bibinfo {author} {\bibfnamefont {M.}~\bibnamefont {Tabak}}, \bibinfo {author} {\bibfnamefont {H.}~\bibnamefont {Chen}}, \bibinfo {author} {\bibfnamefont {H.~K.}\ \bibnamefont {Chung}}, \bibinfo {author} {\bibfnamefont {K.~B.}\ \bibnamefont {Fournier}}, \bibinfo {author} {\bibfnamefont {S.~B.}\ \bibnamefont {Hansen}}, \bibinfo {author} {\bibfnamefont {A.}~\bibnamefont {Kemp}}, \bibinfo {author} {\bibfnamefont {D.~A.}\ \bibnamefont {Liedahl}}, \bibinfo {author} {\bibfnamefont {K.}~\bibnamefont {Widmann}}, \bibinfo {author} {\bibfnamefont {S.~C.}\ \bibnamefont {Wilks}}, \bibinfo {author} {\bibfnamefont {W.}~\bibnamefont {Rozmus}},\ and\ \bibinfo {author} {\bibfnamefont {M.}~\bibnamefont {Sherlock}},\ }\bibfield  {title} {\bibinfo {title} {{Absorption of short laser pulses on solid targets in the
  ultrarelativistic regime}},\ }\href {https://doi.org/10.1103/PhysRevLett.100.085004} {\bibfield  {journal} {\bibinfo  {journal} {Physical Review Letters}\ }\textbf {\bibinfo {volume} {100}},\ \bibinfo {pages} {6} (\bibinfo {year} {2008})}\BibitemShut {NoStop}%
\bibitem [{\citenamefont {Hatchett}\ \emph {et~al.}(2000)\citenamefont {Hatchett}, \citenamefont {Brown}, \citenamefont {Cowan}, \citenamefont {Henry}, \citenamefont {Johnson}, \citenamefont {Key}, \citenamefont {Koch}, \citenamefont {Langdon}, \citenamefont {Lasinski}, \citenamefont {Lee}, \citenamefont {Mackinnon}, \citenamefont {Pennington}, \citenamefont {Perry}, \citenamefont {Phillips}, \citenamefont {Roth}, \citenamefont {Sangster}, \citenamefont {Singh}, \citenamefont {Snavely}, \citenamefont {Stoyer}, \citenamefont {Wilks},\ and\ \citenamefont {Yasuike}}]{Hatchett2000ElectronTargets}%
  \BibitemOpen
  \bibfield  {author} {\bibinfo {author} {\bibfnamefont {S.~P.}\ \bibnamefont {Hatchett}}, \bibinfo {author} {\bibfnamefont {C.~G.}\ \bibnamefont {Brown}}, \bibinfo {author} {\bibfnamefont {T.~E.}\ \bibnamefont {Cowan}}, \bibinfo {author} {\bibfnamefont {E.~A.}\ \bibnamefont {Henry}}, \bibinfo {author} {\bibfnamefont {J.~S.}\ \bibnamefont {Johnson}}, \bibinfo {author} {\bibfnamefont {M.~H.}\ \bibnamefont {Key}}, \bibinfo {author} {\bibfnamefont {J.~A.}\ \bibnamefont {Koch}}, \bibinfo {author} {\bibfnamefont {A.~B.}\ \bibnamefont {Langdon}}, \bibinfo {author} {\bibfnamefont {B.~F.}\ \bibnamefont {Lasinski}}, \bibinfo {author} {\bibfnamefont {R.~W.}\ \bibnamefont {Lee}}, \bibinfo {author} {\bibfnamefont {A.~J.}\ \bibnamefont {Mackinnon}}, \bibinfo {author} {\bibfnamefont {D.~M.}\ \bibnamefont {Pennington}}, \bibinfo {author} {\bibfnamefont {M.~D.}\ \bibnamefont {Perry}}, \bibinfo {author} {\bibfnamefont {T.~W.}\ \bibnamefont {Phillips}}, \bibinfo {author} {\bibfnamefont {M.}~\bibnamefont {Roth}}, \bibinfo
  {author} {\bibfnamefont {T.~C.}\ \bibnamefont {Sangster}}, \bibinfo {author} {\bibfnamefont {M.~S.}\ \bibnamefont {Singh}}, \bibinfo {author} {\bibfnamefont {R.~A.}\ \bibnamefont {Snavely}}, \bibinfo {author} {\bibfnamefont {M.~A.}\ \bibnamefont {Stoyer}}, \bibinfo {author} {\bibfnamefont {S.~C.}\ \bibnamefont {Wilks}},\ and\ \bibinfo {author} {\bibfnamefont {K.}~\bibnamefont {Yasuike}},\ }\bibfield  {title} {\bibinfo {title} {{Electron, photon, and ion beams from the relativistic interaction of Petawatt laser pulses with solid targets}},\ }\href {https://doi.org/10.1063/1.874030} {\bibfield  {journal} {\bibinfo  {journal} {Physics of Plasmas}\ }\textbf {\bibinfo {volume} {7}},\ \bibinfo {pages} {2076} (\bibinfo {year} {2000})}\BibitemShut {NoStop}%
\bibitem [{\citenamefont {Krainara}\ \emph {et~al.}(2018)\citenamefont {Krainara}, \citenamefont {Chatani}, \citenamefont {Zen}, \citenamefont {Kii},\ and\ \citenamefont {Ohgaki}}]{Krainara2018StudySource}%
  \BibitemOpen
  \bibfield  {author} {\bibinfo {author} {\bibfnamefont {S.}~\bibnamefont {Krainara}}, \bibinfo {author} {\bibfnamefont {S.}~\bibnamefont {Chatani}}, \bibinfo {author} {\bibfnamefont {H.}~\bibnamefont {Zen}}, \bibinfo {author} {\bibfnamefont {T.}~\bibnamefont {Kii}},\ and\ \bibinfo {author} {\bibfnamefont {H.}~\bibnamefont {Ohgaki}},\ }\bibfield  {title} {\bibinfo {title} {{Study of the saturation of radiation energy caused by the space charge effect in a compact THz coherent radiation source}},\ }\bibfield  {journal} {\bibinfo  {journal} {Journal of Physics: Conference Series}\ }\textbf {\bibinfo {volume} {1067}},\ \href {https://doi.org/10.1088/1742-6596/1067/3/032022} {10.1088/1742-6596/1067/3/032022} (\bibinfo {year} {2018})\BibitemShut {NoStop}%
\bibitem [{\citenamefont {Liao}\ \emph {et~al.}(2016)\citenamefont {Liao}, \citenamefont {Li}, \citenamefont {Zhang}, \citenamefont {Liu}, \citenamefont {Ge}, \citenamefont {Yang}, \citenamefont {Wei}, \citenamefont {Yuan}, \citenamefont {Deng}, \citenamefont {Zhu}, \citenamefont {Zhang}, \citenamefont {Wang}, \citenamefont {Sheng}, \citenamefont {Chen}, \citenamefont {Lu}, \citenamefont {Ma}, \citenamefont {Wang},\ and\ \citenamefont {Zhang}}]{Liao2016}%
  \BibitemOpen
  \bibfield  {author} {\bibinfo {author} {\bibfnamefont {G.~Q.}\ \bibnamefont {Liao}}, \bibinfo {author} {\bibfnamefont {Y.~T.}\ \bibnamefont {Li}}, \bibinfo {author} {\bibfnamefont {Y.~H.}\ \bibnamefont {Zhang}}, \bibinfo {author} {\bibfnamefont {H.}~\bibnamefont {Liu}}, \bibinfo {author} {\bibfnamefont {X.~L.}\ \bibnamefont {Ge}}, \bibinfo {author} {\bibfnamefont {S.}~\bibnamefont {Yang}}, \bibinfo {author} {\bibfnamefont {W.~Q.}\ \bibnamefont {Wei}}, \bibinfo {author} {\bibfnamefont {X.~H.}\ \bibnamefont {Yuan}}, \bibinfo {author} {\bibfnamefont {Y.~Q.}\ \bibnamefont {Deng}}, \bibinfo {author} {\bibfnamefont {B.~J.}\ \bibnamefont {Zhu}}, \bibinfo {author} {\bibfnamefont {Z.}~\bibnamefont {Zhang}}, \bibinfo {author} {\bibfnamefont {W.~M.}\ \bibnamefont {Wang}}, \bibinfo {author} {\bibfnamefont {Z.~M.}\ \bibnamefont {Sheng}}, \bibinfo {author} {\bibfnamefont {L.~M.}\ \bibnamefont {Chen}}, \bibinfo {author} {\bibfnamefont {X.}~\bibnamefont {Lu}}, \bibinfo {author} {\bibfnamefont {J.~L.}\ \bibnamefont {Ma}},
  \bibinfo {author} {\bibfnamefont {X.}~\bibnamefont {Wang}},\ and\ \bibinfo {author} {\bibfnamefont {J.}~\bibnamefont {Zhang}},\ }\bibfield  {title} {\bibinfo {title} {{Demonstration of Coherent Terahertz Transition Radiation from Relativistic Laser-Solid Interactions}},\ }\href {https://doi.org/10.1103/PhysRevLett.116.205003} {\bibfield  {journal} {\bibinfo  {journal} {Physical Review Letters}\ }\textbf {\bibinfo {volume} {116}},\ \bibinfo {pages} {1} (\bibinfo {year} {2016})}\BibitemShut {NoStop}%
\bibitem [{\citenamefont {Gopal}\ \emph {et~al.}(2013)\citenamefont {Gopal}, \citenamefont {Singh}, \citenamefont {Herzer}, \citenamefont {Reinhard}, \citenamefont {Schmidt}, \citenamefont {Dillner}, \citenamefont {May}, \citenamefont {Meyer}, \citenamefont {Ziegler},\ and\ \citenamefont {Paulus}}]{Gopal2013CharacterizationInteraction}%
  \BibitemOpen
  \bibfield  {author} {\bibinfo {author} {\bibfnamefont {A.}~\bibnamefont {Gopal}}, \bibinfo {author} {\bibfnamefont {P.}~\bibnamefont {Singh}}, \bibinfo {author} {\bibfnamefont {S.}~\bibnamefont {Herzer}}, \bibinfo {author} {\bibfnamefont {A.}~\bibnamefont {Reinhard}}, \bibinfo {author} {\bibfnamefont {A.}~\bibnamefont {Schmidt}}, \bibinfo {author} {\bibfnamefont {U.}~\bibnamefont {Dillner}}, \bibinfo {author} {\bibfnamefont {T.}~\bibnamefont {May}}, \bibinfo {author} {\bibfnamefont {H.-G.}\ \bibnamefont {Meyer}}, \bibinfo {author} {\bibfnamefont {W.}~\bibnamefont {Ziegler}},\ and\ \bibinfo {author} {\bibfnamefont {G.~G.}\ \bibnamefont {Paulus}},\ }\bibfield  {title} {\bibinfo {title} {{Characterization of 700 {$\mu$}J T rays generated during high-power laser solid interaction}},\ }\href {https://doi.org/10.1364/ol.38.004705} {\bibfield  {journal} {\bibinfo  {journal} {Optics Letters}\ }\textbf {\bibinfo {volume} {38}},\ \bibinfo {pages} {4705} (\bibinfo {year} {2013})}\BibitemShut {NoStop}%
\bibitem [{\citenamefont {Casalbuoni}\ \emph {et~al.}(2009)\citenamefont {Casalbuoni}, \citenamefont {Schmidt}, \citenamefont {Schm{\"{u}}ser}, \citenamefont {Arsov},\ and\ \citenamefont {Wesch}}]{Casalbuoni2009}%
  \BibitemOpen
  \bibfield  {author} {\bibinfo {author} {\bibfnamefont {S.}~\bibnamefont {Casalbuoni}}, \bibinfo {author} {\bibfnamefont {B.}~\bibnamefont {Schmidt}}, \bibinfo {author} {\bibfnamefont {P.}~\bibnamefont {Schm{\"{u}}ser}}, \bibinfo {author} {\bibfnamefont {V.}~\bibnamefont {Arsov}},\ and\ \bibinfo {author} {\bibfnamefont {S.}~\bibnamefont {Wesch}},\ }\bibfield  {title} {\bibinfo {title} {{Ultrabroadband terahertz source and beamline based on coherent transition radiation}},\ }\href {https://doi.org/10.1103/PhysRevSTAB.12.030705} {\bibfield  {journal} {\bibinfo  {journal} {Physical Review Special Topics - Accelerators and Beams}\ }\textbf {\bibinfo {volume} {12}},\ \bibinfo {pages} {1} (\bibinfo {year} {2009})}\BibitemShut {NoStop}%
\bibitem [{\citenamefont {Di~Mitri}\ \emph {et~al.}(2018)\citenamefont {Di~Mitri}, \citenamefont {Perucchi}, \citenamefont {Adhlakha}, \citenamefont {Di~Pietro}, \citenamefont {Nicastro}, \citenamefont {Roussel}, \citenamefont {Spampinati}, \citenamefont {Veronese}, \citenamefont {Allaria}, \citenamefont {Badano}, \citenamefont {Cudin}, \citenamefont {De~Ninno}, \citenamefont {Diviacco}, \citenamefont {Gaio}, \citenamefont {Gauthier}, \citenamefont {Giannessi}, \citenamefont {Lupi}, \citenamefont {Penco}, \citenamefont {Piccirilli}, \citenamefont {Rebernik}, \citenamefont {Spezzani},\ and\ \citenamefont {Trov{\`{o}}}}]{DiMitri2018CoherentWakefield}%
  \BibitemOpen
  \bibfield  {author} {\bibinfo {author} {\bibfnamefont {S.}~\bibnamefont {Di~Mitri}}, \bibinfo {author} {\bibfnamefont {A.}~\bibnamefont {Perucchi}}, \bibinfo {author} {\bibfnamefont {N.}~\bibnamefont {Adhlakha}}, \bibinfo {author} {\bibfnamefont {P.}~\bibnamefont {Di~Pietro}}, \bibinfo {author} {\bibfnamefont {S.}~\bibnamefont {Nicastro}}, \bibinfo {author} {\bibfnamefont {E.}~\bibnamefont {Roussel}}, \bibinfo {author} {\bibfnamefont {S.}~\bibnamefont {Spampinati}}, \bibinfo {author} {\bibfnamefont {M.}~\bibnamefont {Veronese}}, \bibinfo {author} {\bibfnamefont {E.}~\bibnamefont {Allaria}}, \bibinfo {author} {\bibfnamefont {L.}~\bibnamefont {Badano}}, \bibinfo {author} {\bibfnamefont {I.}~\bibnamefont {Cudin}}, \bibinfo {author} {\bibfnamefont {G.}~\bibnamefont {De~Ninno}}, \bibinfo {author} {\bibfnamefont {B.}~\bibnamefont {Diviacco}}, \bibinfo {author} {\bibfnamefont {G.}~\bibnamefont {Gaio}}, \bibinfo {author} {\bibfnamefont {D.}~\bibnamefont {Gauthier}}, \bibinfo {author} {\bibfnamefont
  {L.}~\bibnamefont {Giannessi}}, \bibinfo {author} {\bibfnamefont {S.}~\bibnamefont {Lupi}}, \bibinfo {author} {\bibfnamefont {G.}~\bibnamefont {Penco}}, \bibinfo {author} {\bibfnamefont {F.}~\bibnamefont {Piccirilli}}, \bibinfo {author} {\bibfnamefont {P.}~\bibnamefont {Rebernik}}, \bibinfo {author} {\bibfnamefont {C.}~\bibnamefont {Spezzani}},\ and\ \bibinfo {author} {\bibfnamefont {M.}~\bibnamefont {Trov{\`{o}}}},\ }\bibfield  {title} {\bibinfo {title} {{Coherent THz Emission Enhanced by Coherent Synchrotron Radiation Wakefield}},\ }\href {https://doi.org/10.1038/s41598-018-30125-1} {\bibfield  {journal} {\bibinfo  {journal} {Scientific Reports}\ }\textbf {\bibinfo {volume} {8}},\ \bibinfo {pages} {1} (\bibinfo {year} {2018})}\BibitemShut {NoStop}%
\bibitem [{\citenamefont {Boscolo}\ \emph {et~al.}(2009)\citenamefont {Boscolo}, \citenamefont {Castellano}, \citenamefont {Chiadroni}, \citenamefont {Ferrario}, \citenamefont {Calvani}, \citenamefont {Lupi}, \citenamefont {Nucara}, \citenamefont {Vergata}, \citenamefont {Petrillo},\ and\ \citenamefont {Trieste}}]{Boscolo2009DESIGNLINAC}%
  \BibitemOpen
  \bibfield  {author} {\bibinfo {author} {\bibfnamefont {M.}~\bibnamefont {Boscolo}}, \bibinfo {author} {\bibfnamefont {M.}~\bibnamefont {Castellano}}, \bibinfo {author} {\bibfnamefont {E.}~\bibnamefont {Chiadroni}}, \bibinfo {author} {\bibfnamefont {M.}~\bibnamefont {Ferrario}}, \bibinfo {author} {\bibfnamefont {P.}~\bibnamefont {Calvani}}, \bibinfo {author} {\bibfnamefont {S.}~\bibnamefont {Lupi}}, \bibinfo {author} {\bibfnamefont {A.}~\bibnamefont {Nucara}}, \bibinfo {author} {\bibfnamefont {T.}~\bibnamefont {Vergata}}, \bibinfo {author} {\bibfnamefont {V.}~\bibnamefont {Petrillo}},\ and\ \bibinfo {author} {\bibfnamefont {A.~P.~E.}\ \bibnamefont {Trieste}},\ }\bibfield  {title} {\bibinfo {title} {{DESIGN STUDY OF A DEDICATED BEAMLINE FOR TH z RADIATION GENERATION AT THE SPARC LINAC}},\ }\href@noop {} {\bibfield  {journal} {\bibinfo  {journal} {Proceedings of PAC09}\ }\textbf {\bibinfo {volume} {1}},\ \bibinfo {pages} {1168} (\bibinfo {year} {2009})}\BibitemShut {NoStop}%
\end{thebibliography}%

\end{document}